# Photon bunching in high-harmonic emission controlled by quantum light.

*Samuel Lemieux,[1] Sohail A. Jalil,[1] David Purschke,[1] Neda Boroumand,[2] David Villeneuve,[1] Andrei Naumov,[1] Thomas Brabec,[2] Giulio Vampa[1]*

[1] *Joint Attosecond Science Laboratory, National Research Council of Canada and University of Ottawa, Ottawa, Ontario, K1N 0R6, Canada*

[2] *Department of Physics, University of Ottawa, 150 Louis Pasteur, Ottawa, Ontario K1N 6N5, Canada*

*gvampa@uottawa.ca

**Attosecond spectroscopy comprises several techniques to probe matter through electrons and photons[1,2]. One frontier of attosecond methods is to reveal complex phenomena arising from quantum-mechanical correlations in the matter system[3,4], in the photon fields[5-8] and among them. Recent theories have laid the groundwork for understanding how quantum-optical properties affect high-field photonics, such as strong-field ionization and acceleration of electrons in quantum-optical fields[9], and how entanglement between the field modes arises during the interaction[6]. Here we demonstrate a new experimental approach that transduces some properties of a quantum-optical state through a strong-field nonlinearity. We perturb high-harmonic emission from a semiconductor with a bright squeezed vacuum field resulting in the emission of sidebands of the high-harmonics with super-Poissonian statistics, indicating that the emitted photons are *bunched*. Our results suggest that perturbing strong-field dynamics with quantum-optical states is a viable way to coherently control the generation of these states at short wavelengths, such as extreme ultraviolet or soft X-rays. Quantum correlations will be instrumental to advance attosecond spectroscopy and imaging beyond the classical limits.**

Quantum-optical states of light are finding numerous applications in fields such as computing, sensing and metrology[10], spectroscopy[11,12] and imaging[13]. Most of the technology relies on the manipulation and readout of photons at infrared or visible frequencies. Accessing quantum light at shorter wavelengths, such as extreme ultraviolet (XUV) or X-rays[14,15], may be quite beneficial: phase sensitivity increases, a feature particularly appealing for sensing applications, while detector noise decreases because the photon energy is much greater than thermal fluctuations. Yet, investigation of quantum-optical states at short wavelengths is still in its infancy[16,17]. Short-wavelength radiation can be produced in a laboratory via high-harmonic generation[18], an extremely nonlinear optical process that creates a veritable attosecond-lived oscillating quantum antenna composed of laser-accelerated electron-hole pairs in matter that radiates light at odd-multiples of the driving laser frequency[19,20]. Exploring high-harmonic emission from a quantum-optical perspective is poised to reveal new insights into strong light-matter interactions[21,6-9] and unlock new capabilities, such as the generation of cat or kitten states[5], squeezed XUV light[8], and entangled XUV photons.

Here we investigate, experimentally, how strongly driven electron-hole pairs responsible for high-harmonic emission from a semiconductor (ZnO) react to a quantum-optical perturbation, a Bright Squeezed Vacuum (BSV) beam. The BSV beam creates sidebands of the unperturbed odd-harmonic spectrum that exhibit shot-to-shot power fluctuations with super-Poissonian statistics, indicative of *photon bunching* and of the generation of a non-coherent state at the short high-harmonic wavelength. Furthermore, our simple theoretical model predicts that the variance of the field quadratures of the sidebands can be coherently controlled with the relative phase between BSV and the unperturbed spectrum, paving the way for the controlled generation of non-coherent quantum-optical attosecond states.

To perturb electron trajectories leading to high-harmonic emission, we mix a BSV beam at a wavelength of 1600 nm with an intense coherent mid-infrared beam with a wavelength of 3350 nm, inside a ZnO (0001) single crystal (Fig. 1**a**). The BSV beam is obtained by Spontaneous Parametric Down Conversion (SPDC) of a 50-fs, 800-nm laser in two successive BBO crystals, which filters the number of spatial and spectral modes to nearly 1 [22] (see Methods). The statistics of the measured pulse energy of each shot in these two beams confirm that the mid-infrared pump is in a coherent state, with a normalized second-order correlation function $g^{(2)}(0) = 1.05 \pm 0.02$ (Fig. 1**b**), whereas the BSV beam exhibits a heavy-tail distribution with $g^{(2)}(0) = 2.24 \pm 0.34$ (Fig. 1**c**) typical of femtosecond BSV[23]. At an intensity of ~ 0.5 TW/cm$^2$, the mid-infrared driver generates only odd-order harmonics down to ~ 210 nm wavelength (5.9 eV photon energy), see spectrum in Fig. 1**d**. The addition of the BSV beam generates sidebands at nearly the even-order harmonics of the coherent pump (Fig. 1**e**). The emission of even-order sidebands indicates that the perturbation to the high-harmonic dipole $\sigma$ obeys the following time-translation symmetry: $\sigma\left(t + \frac{\pi}{\omega}\right) = -\sigma(t)$ [24, 25], e.g. it changes sign between two adjacent laser half-cycles ($\omega$ is the frequency of the driving field). Thus, even though the BSV field is not well defined at any instant, it retains the time-translation symmetry of a coherent field. In other words, it is first-order coherent[26] and repeats identical every cycle within the coherence length, with amplitude fluctuations only arising among different temporal modes (pulses).

The photon statistics of the generated ultraviolet light are strikingly different between sidebands and harmonics: whereas the harmonics exhibit statistics resembling that of the coherent pump (Fig. 1**f-i**), the sidebands inherit the heavy-tailed distribution of the BSV beam

(Fig. 1**l-o**). For comparison, the Poisson distribution with the mean number of photons in the 8$^{th}$ sideband is given in Fig. 1**l**, orange line. This split between harmonics and sidebands is even more apparent when comparing the $g^{(2)}(0)$. Figure 2 reports the $g^{(2)}(0)$ for 20 blocks of the data for each harmonic and sideband, as a function of the measured average number of photons/shot. Whereas the harmonics maintain a $g^{(2)}(0) < 1.5$, the sidebands exhibit $g^{(2)}(0) > 1.6$, with a trend of increasing $g^{(2)}(0)$ for decreasing sideband order, or increasing average photon number. For comparison, the BSV exhibits a $g^{(2)}(0) \sim 2.2$ (gray shaded area in Fig. 2). Thus, the sidebands exhibit super-Poissonian statistics with fluctuations that can be either smaller or larger than those of the BSV. Since the emission occurs within the femtosecond pulse duration, $g^{(2)}(\tau>0) = 1$, thus $g^{(2)}(0) > g^{(2)}(\tau>0)$, i.e. the sidebands are likely *photon bunched*[26].

Only ~10 nJ of BSV are needed to generate a broad comb of non-Poissonian light across the visible and deep-ultraviolet spectral regions (ZnO is known to generate high-harmonics up to 10 eV photon energies[27], thus the generation of sidebands in the vacuum-ultraviolet spectral region is plausible). The generation of sidebands across the whole high-harmonic spectrum indicates that high-harmonic generation gates the interaction with the BSV pulse to a fraction of the coherent mid-infrared cycle, periodically every half cycle. The duration of the gate depends on the duration of the perturbed electron-hole pair trajectory[25], thus it can be as short as attoseconds. A temporal characterization of the sidebands is likely to yield a train of attosecond to few-femtosecond bursts of non-classical light. In addition, each sideband has a spectral bandwidth comparable to the high harmonics, evidence of further temporal gating within the mid-infrared pulse envelope. In contrast, direct up-conversion of the BSV beam, which has a 50 nm bandwidth at 1600 nm, results in narrow low-order harmonics (Extended Data 1), which are significantly weaker than the sidebands. Thus, our approach leverages the strength of the coherent mid-infrared driver to generate brighter non-coherent states at short wavelengths. Gating also occurs in space, which results in a filtering of the number of spatial sideband modes. Thus, *in-situ* control of high-harmonic generation with non-classical fields may be an effective way to study the quantum-optical nature of strong-field interactions with readily available quantum-optical fields. For example, to measure the *photon-statistics force* theorized in Ref. 9. Our results indicate that strong-field excitation, acceleration and recombination of electron-hole pairs provide a sub-cycle gate to control quantum-optical states, which can be harnessed to further our understanding of time-resolved quantum electrodynamics[28,29]. The same method can potentially be used to measure quantum-optical states, provided their interaction with the high-harmonic generation process is well understood.

The sideband power varies linearly with BSV power (Extended Data 2), as expected from a wavemixing picture of high-harmonic generation perturbed with classical fields[30,31]. Linear scaling in the perturbing field is interpreted as arising from 1-photon addition/subtraction channels from the sub-laser-cycle high-harmonic spectrum. Based on this understanding, we model the generation of sidebands with the following interaction Hamiltonian for the radiation field:

$$H_{int} = i\hbar\gamma\left(\epsilon_1 \hat{a}_0 \hat{a}_{sb}^\dagger + \epsilon_2 \hat{a}_0^\dagger \hat{a}_{sb}^\dagger + h.c.\right) \qquad (1)$$

Where the first two terms represent emission of a sideband photon ($\hat{a}_{sb}^\dagger$) by absorption of one sideband photon ($\hat{a}_0$) from a lower-energy high-harmonic field ($\epsilon_1$), equivalent to a *sum-frequency* process, and emission of one perturbing photon ($\hat{a}_0^\dagger$) from a higher-energy high-harmonic field ($\epsilon_2$), equivalent to a *difference-frequency* process (see Fig. 3**a**). The unperturbed high-harmonic states are considered coherent and therefore treated classically, thus $\epsilon_1, \epsilon_2 \in \mathbb{C}$. The complex $\epsilon_1$ and $\epsilon_2$ account for the varying spectral phases $\theta_1$ and $\theta_2$ of the high harmonics (*attochirp*[20]). The perturbing field is assumed to be a squeezed vacuum state with squeezing parameter $\xi = |r|e^{i\phi}$ [26]. The mean sideband photon number, $\langle n \rangle$, and the variance of the two field quadratures, $\langle (\Delta X)^2 \rangle$ and $\langle (\Delta P)^2 \rangle$, are plotted in Fig. 3**b-d** respectively, as a function of $\phi$ and $\theta_1 - \theta_2$. We find that $\langle n \rangle$ oscillates because of interference between the degenerate sum- and difference-frequency pathways. This behavior is analogous to that measured from classical fields[24,25]. The quadratures, on the other hand, reveal a more nuanced modulation: depending on the relative phase between the initial high-harmonic states and the squeezed perturbation, the quadratures can be equal (Fig. 3**f**), leading to the generation of *super-thermal* sidebands[32,33], or unequal, resulting in *squashed* states[34]. In the case where $\theta_1 - \theta_2 = 0$ (Fig. 3**f**), i.e. the harmonics are in phase, one quadrature modulates with $\phi$ whereas the other retains minimum uncertainty ($4\langle (\Delta P)^2 \rangle = 1$), resulting in a maximally squashed state. Squashed and super-thermal states have found use in quantum-enhanced imaging[35] and quantum computation[34]. Additionally, the model predicts that the spent perturbing field and the emitted sideband are correlated since the latter requires absorption of a squeezed photon. A projective measurement on a sideband photon, therefore, may result in the generation of a coherently controlled cat or kitten state in the perturbing squeezed beam[5].

Is it commonly understood that the generation of squeezed (e.g. below shot noise) sum-frequency is possible when nearly all photons of the initial lower-frequency squeezed beam are upconverted[36]. This is also what our model predicts when the perturbing field is depleted (Fig.

3f, yellow line). This is because squeezed vacuum photons are correlated in pairs. Preserving squeezing at the sum-frequency requires up-conversion of both correlated photons in a pair, a condition that is only reached when most photons are converted if the conversion process doesn't preferentially select pairs. The quantum correlation between pairs of photons results in time-dependent fluctuations of the electric field that alternate regions of sub-shot noise (the *squeezed* quadrature) with regions of large excess noise (the *antisqueezed* quadrature). However, in the experiment, we estimate that ~ 10 photons/shot are emitted from the crystal at the $12^{th}$ sideband, when the average pulse energy of the BSV is set to ~ 7 nJ, corresponding to ~ 5 x $10^{10}$ photons/shot. Thus, approximately 1 sideband photon is emitted per $10^9$ BSV photons. This conversion efficiency is used for the modelling. With this low conversion efficiency, squeezing gives way to the emission of squashed or thermal sidebands.

In summary, we experimentally demonstrated that high-harmonic generation can transduce some aspects of a quantum-optical perturbation, bright squeezed vacuum, to high-order sidebands. Our results elicit questions as to how quantum-optical fields transfer *quantumness* to the accelerated electron-hole pairs responsible for high-harmonic emission in crystals. The same method can likely be applied to the generation of unusual coherently-controlled attosecond currents[37,38] and THz radiation[39]. Finally, our method paves the way for coherently controlled emission of non-classical states with high photon energy, with potential benefits for quantum-enhanced metrology at short wavelengths, possibly even in the extreme ultraviolet, and in the attosecond temporal domain.

**Methods**

**Experimental layout.** The sketch of the experimental layout is shown in Extended Data 3. A femtosecond amplified Ti:Sapphire laser system (Coherent Legend) delivers 50 fs pulses with 1.6 mJ of energy, at a repetition rate of 1 kHz. 1.3 mJ pump a super-fluorescence seeded Optical Parametric Amplifier (LightConversion TOPAS). The signal and idler beams are set to wavelengths of 1.29 µm and 2.1 µm, respectively. They are spatially separated and recombined on a Type II AGS crystal tuned for difference frequency generation at a wavelength of 3.35 µm. A long-pass filter with a cut-on wavelength of 2.4 µm (Edmund #68-653) removes signal and idler beams. About 1 µJ of energy is measured in the Mid-infrared beam after this filter. The mid-infrared beam, the coherent pump in the experiment, is telescoped 1:2 with $CaF_2$ lenses and focused on a 200 µm-thick ZnO(0001) single crystal with an off-axis parabola with f = 50 mm, generating odd-order high harmonics. The size of the high-harmonic beam at the ZnO exit surface is 40 µm ($1/e^2$ diameter), measured with the drop of harmonic power as a sharp Au electrode deposited on the ZnO surface is scanned over the beam. Using the high-harmonic spot size, the estimated mid-infrared pulse duration of 80 fs, and the energy incident on ZnO (300 nJ), we estimate the mid-infrared intensity to be comparable or less than 0.3 $TW/cm^2$. The harmonics are focused on the input slit of a commercial UV-VIS spectrometer (Princeton Instruments Isoplane 320, equipped with a PI-MAX4 intensified camera with SG-UV photocathode).

The remaining 300 uJ from the laser system are utilized for generation of a BSV beam in Type I geometry in two 2 mm-thick, uncoated, BBO crystals in tandem. The pump beam is telescoped down to ~ 1 mm size before entering the BBO crystals, and a combination of wavelength and transmission polarizer serve as power throttle. The serial combination of BBOs effectively filters a smaller selection of spatial (and frequency) modes emanating from the first crystal, as demonstrated in [22]. The BSV spectrum is subsequently filtered to a 50 nm bandwidth centered about 1.6 μm (Edmund #87-872), and spurious visible light is further suppressed with an RG1000 filter. The BSV beam at the BBO output is re-imaged on the ZnO crystal with suitable demagnification to a spot size of 70 – 100 μm ($1/e^2$ diameter), as measured with an infrared objective lens (10x Mitutoyo Plan Apo NIR Infinity Corrected) and an InGaAs camera (Xenics Bobcat+ 320). Collinear combination of BSV and mid-infrared beams is performed with a custom dielectric mirror (S1 HR 1600-2000 nm HT 3200 – 4200 nm, S2 uncoated, $CaF_2$ substrate, Laseroptik GmbH). Spatial overlap at the ZnO crystal is measured with the same imaging system utilized to measure the BSV focus size. Both beams have parallel polarizations.

For measuring the single-shot power of the sidebands and harmonics, the ultraviolet beam is sent to a custom-made spectrometer composed of a 90 deg, $CaF_2$ prism, a $CaF_2$ lens (f = 250 mm), and a photomultiplier tube (PMT, Hamamatsu H9305-01). The $g^{(2)}(0)$ values are calculated as $g^{(2)}(0) = \langle n^2 \rangle / \langle n \rangle^2 - 1/\langle n \rangle$ with $n$ being the number of photons/shot measured by the single detector. This expression is valid for a single detector as long as the number of spatio-temporal modes is close to 1 [40], which is the case in our experiment. The reflection of the mid-infrared pump off the surface of a $CaF_2$ window is measured with a PbSe amplified photodiode (Thorlabs PDA20H). The reflection of the BSV off the RG1000 filter is measured with an InGaAs photodiode (Thorlabs DET10D). Single shots of all three beams (PMT, BSV and mid-infrared) are acquired simultaneously with a boxcar integrator and a 12-bit DAQ.

The acquired signal is converted to average number of photons/shot as follows:

$$S_{out}\{ph/shot\} = S_{in}\{ADU\} \times \frac{ADC\ range\ (V)}{4096} \times \frac{Boxcar\ sens\ (V_{in}/V_{out})}{1\ M\Omega \times e \times PMT\ gain\ \times 1\ kHz} \times \frac{1}{QE}$$

where *ADC range* = 20 V, *Boxcar sens* = 0.02 V/V is the boxcar sensitivity, 1 MΩ accounts for the input impedance of the boxcar, $e = 1.6 \times 10^{-19}$ C is the electron charge, *PMT gain* is the gain of the PMT ($\sim 10^7$), and *QE* is the quantum efficiency of the PMT, which is ~ 50-80% across the measured spectral range.

**Theoretical model.** The initial photon state is considered to be $|\varphi_0(0)\rangle = |\xi\rangle_0 |\epsilon_1\rangle_1 |0\rangle_{sb} |\epsilon_2\rangle_2$, where subscripts 0, 1, 'sb' and 2 refer to the perturbation, the lower-lying harmonic, the sideband and the higher-lying harmonic (see sketch in Fig. 3a). The perturbation is in a squeezed vacuum state with squeezing parameter $\xi = |r|e^{i\phi}$, the harmonics are in coherent states with amplitudes $\epsilon_1 = |\epsilon_1|e^{i\theta_1}$ and $\epsilon_2 = |\epsilon_2|e^{i\theta_2}$, and the sideband is initially in vacuum. The Hamiltonian in Eq. (1) leads to the final state:

$$|\varphi_0(t)\rangle = e^{g\hat{a}_0 \hat{a}_{sb}^\dagger + f \hat{a}_0^\dagger \hat{a}_{sb}^\dagger + h.c.} |r\rangle_0 |0\rangle_{sb} = \hat{S}|r\rangle_0 |0\rangle_{sb},$$

With $g = \gamma t \epsilon_1$ and $f = \gamma t \epsilon_2$, where '$t$' is the interaction time, representing the efficiencies of the sum-frequency and difference-frequency channels, respectively. The field quadratures are given by $\hat{X} = \hat{a} + \hat{a}^\dagger$, and $\hat{P} = -i(\hat{a} - \hat{a}^\dagger)$. The variances $\langle(\Delta X_{sb})^2\rangle$ and $\langle(\Delta P_{sb})^2\rangle$ and the mean photon number $\langle n_{sb}\rangle$ are calculated as: $\langle\varphi_0|(X_{sb} - \langle X_{sb}\rangle)^2|\varphi_0\rangle = 1 + \langle\varphi_0|\hat{a}^2 + \hat{a}^{\dagger 2} + 2\hat{a}^\dagger \hat{a}|\varphi_0\rangle$, $\langle\varphi_0|(P_{sb} - \langle P_{sb}\rangle)^2|\varphi_0\rangle = 1 - \langle\varphi_0|\hat{a}^2 + \hat{a}^{\dagger 2} - 2\hat{a}^\dagger \hat{a}|\varphi_0\rangle$, and $\langle\varphi_0|\hat{a}_{sb}^\dagger \hat{a}_{sb}|\varphi_0\rangle$. We use the Bogoliubov transformation:

$$\hat{S}^\dagger \hat{a}_{sb} \hat{S} = \hat{a}_{sb} \cosh(N_{sb}) + \hat{p}_{sb}^\dagger \frac{\sinh(N_{sb})}{N_{sb}},$$

where $\hat{p}_{sb}^\dagger = g^* \hat{a}_0^\dagger + f^* \hat{a}_0$ and $N_{sb}^2 = |f|^2 - |g|^2$ to calculate the individual terms in the expectation values. These are:

$$\langle\varphi_0|\hat{a}_{sb}^2|\varphi_0\rangle = M_{sb}^2[fg(\sinh^2 r + \cosh^2 r) - (f^2 e^{-i\phi} + g^2 e^{i\phi})\sinh r \cosh r],$$

$$\langle\varphi_0|\hat{a}_{sb}^{\dagger 2}|\varphi_0\rangle = M_{sb}^2[f^*g^*(\sinh^2 r + \cosh^2 r) - (f^{*2} e^{i\phi} + g^{*2} e^{-i\phi})\sinh r \cosh r],$$

$$\langle n_{sb}\rangle = \langle\varphi_0|\hat{a}_{sb}^{\dagger}\hat{a}_{sb}|\varphi_0\rangle = M_{sb}^2 |f \cosh r - g e^{i\phi} \sinh r|^2.$$

With $M_{sb}^2 = \sinh^2 N_{sb}/N_{sb}$. The result for the quadratures is:

$$\Delta_{X^2}^{Y^2} = 1 + 2M_{sb}^2\left(|f\cosh r - g e^{i\phi}\sinh r|^2 \pm \Re[fg](\sinh^2 r + \cosh^2 r)\right.$$

$$\left.\mp \Re[f^2 e^{-i\phi} + g^2 e^{i\phi}]\sinh r \cosh r\right)$$

According to the definition of the quadratures, the vacuum state has $\Delta_{X^2}^{Y^2} = 1$. Figure 3 plots $\Delta_{X^2}^{Y^2}$ and $\langle n_{sb}\rangle$ for the following parameters: $\sinh r = \sqrt{1e11}$, $\gamma t = \pi/4e5$, $|\epsilon_1| = |\epsilon_2| = 1$, which results in a conversion efficiency of 2 x 10$^{-10}$, similar to that measured experimentally. To show the possibility of squeezing in the sum-frequency pathway alone, we set $|\epsilon_2| = 1$, $\gamma t = \pi/2$, $\sinh r = \sqrt{20}$ (yellow line in Fig. 3**f**). Squeezing only occurs when the conversion efficiency from the squeezed perturbation approaches unity, in agreement with Ref. [36]. The difference frequency pathway, instead, always results in excess quadrature noise because it is effectively a two-mode squeezing interaction, which is known to create states with thermal statistics in each mode[26].

**Data Availability:** Data available on request from the authors.

**Acknowledgements:** It is our pleasant duty to acknowledge David Crane and Ryan Kroeker for providing continuing technical support; Ben Sussman, Guillaume Thekkadath, Frédéric Bouchard, Jeff Lundeen, David Reis and Paul Corkum for useful discussions; Jeremy Upham for lending specialized equipment. G. V. and S. A. J. acknowledge financial support from the Joint Center for Extreme Photonics. G .V. acknowledges support from NSERC Discovery grant RGPIN-2021-04286. D. P. acknowledges support from the NSERC Postdoctoral fellowship program. S. L. and G. V. acknowledge support from the NRC Quantum Sensing program, project QSP-103.

**Author contributions:** G. V. conceived the experiment; S. L., S. A. J., D. N. P. performed the experiment with help from N. B. and D. V.; N. B. and T. B. developed the theoretical model; A. N. maintained the infrastructure; A. S., T. B. and G. V. supervised the project. G. V. wrote the manuscript with help from all co-authors.

**Competing interests.** The authors declare no competing interests.

Correspondence and requests for materials should be addressed to G. Vampa (gvampa@uottawa.ca).

**Figures**

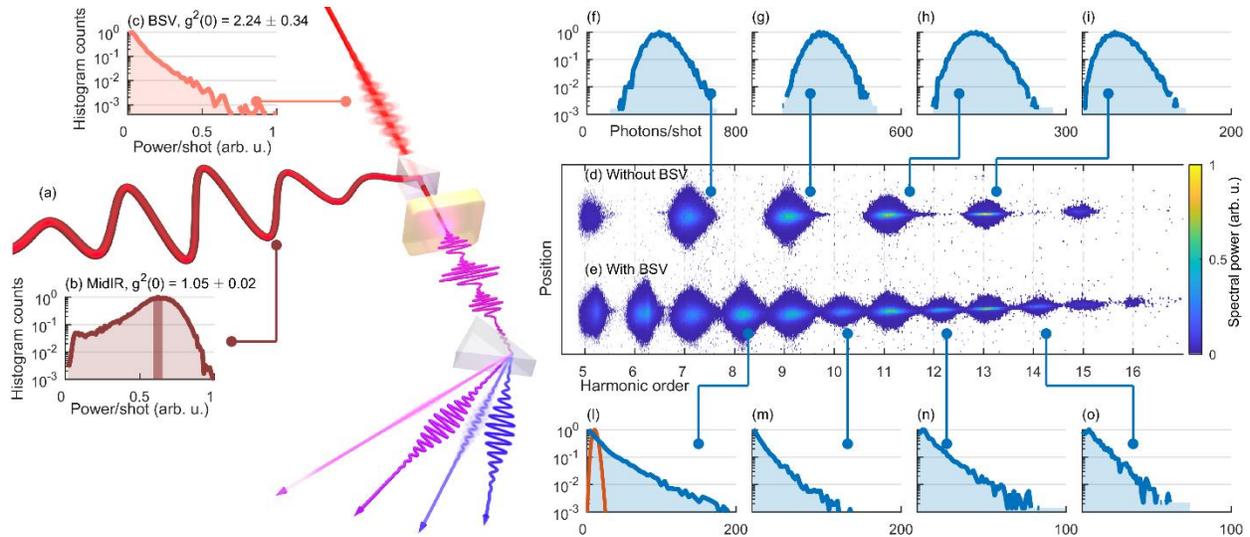

**Figure 1 | Generation of non-Poissonian sidebands. (a)** Intense coherent mid-infrared laser pulses (solid red line, $\lambda = 3350$ nm, $I \sim 0.5$ TW/cm$^2$, $\tau = 80$ fs, u = 300 nJ) are overlapped in space and time with BSV pulses (fuzzy orange line, $\lambda = 1600$ nm, $\Delta\lambda = 50$ nm, u = 30 nJ) inside

a 200 μm-thick ZnO (0001) crystal and generate a comb of high-order harmonics (purple lines). **(b)** Histogram of the pulse energy of the coherent mid-infrared beam. The large non-zero average energy and shape of the distribution results in a $g^{(2)}(0) = 1.05 \pm 0.02$, as expected for a coherent beam. Only pulses withing the red vertical band (20% the standard deviation on either side of the mean) are post-selected for the analysis of the sidebands. The low energy tail stems from fluctuations of the laser system (see Methods). **(c)** Pulse energy statistics of the BSV beam, clearly showing the heavy-tail distribution expected from squeezed vacuum. The calculated $g^{(2)}(0)$ is $2.24 \pm 0.34$. **(d)** The high-harmonic spectrum without BSV perturbation shows that only odd-order harmonics are measured, with statistics **(f-i)** like that of the mid-infrared beam. **(e)** The BSV perturbation generates sidebands of the high-harmonic spectrum at nearly the even-order harmonics. Their statistics closely resemble that of the BSV beam **(l-o)**, clearly deviating from Poisson statistics (the Poisson distribution with the mean number of photons measured for the $8^{th}$ sideband is given in panel **l**, red line). Statistics for the harmonics and sidebands are given in number of photons/shot.

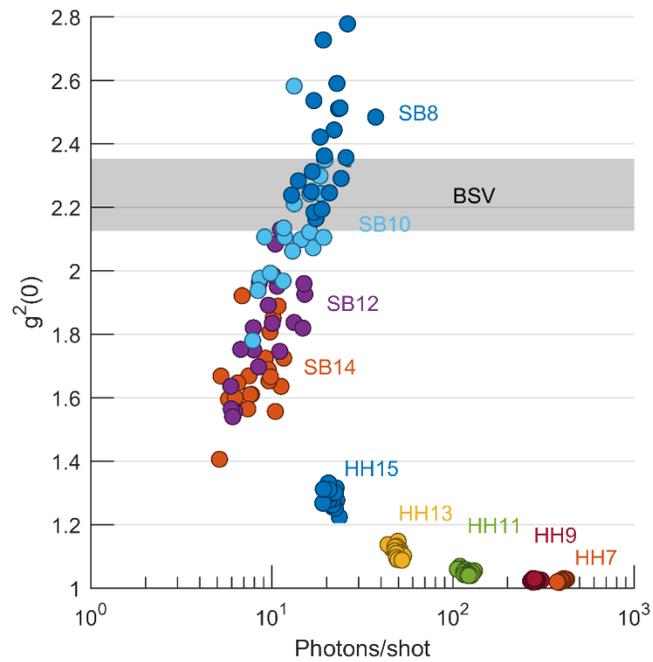

**Figure 2 | Measurement of $g^{(2)}(0)$.** $g^{(2)}(0)$ of the high harmonics is less than 1.4, and tends to 1 for the lowest harmonics (as expected of coherent beams), while for the sidebands $g^{(2)}(0) > 1.6$, and can even exceed that of the BSV beam (gray shaded area). The data set comprises between 15,000 and 30,000 shots, divided in 20 blocks (one data point per block).

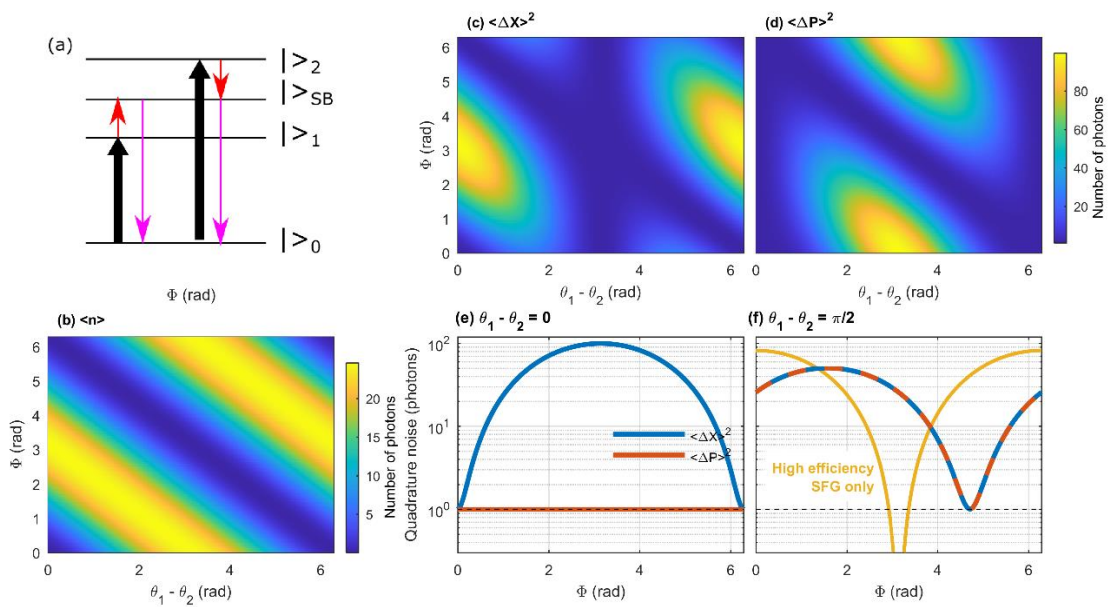

**Figure 3 | Wave-mixing model. (a)** The model accounts for emission of a sideband photon (purple arrows) by absorption of a photon from a lower- and upper-lying high-harmonic states ($|\ \rangle_1$ and $|\ \rangle_2$, respectively, black arrows), and the absorption and emission of a perturbing photon (red arrows), respectively for the two channels. **(b)** The average number of sideband photons $\langle n \rangle$ oscillates with the BSV phase $\phi$ and the relative phase between the un-perturbed high-harmonic states, $\theta_1 - \theta_2$, in agreement with classical models of coherent control. **(c,d)** The $X$ and $P$ quadratures of the sideband field also modulate. Depending on $\theta_1 - \theta_2$ the quadratures can be different from one another (panel **e**, case when the $P$ quadrature retains minimum uncertainty), or equal (panel **f**), leading to either squashed or super-thermal states. The amplitude of the quadratures can be controlled by $\phi$. Squeezing can be achieved in the sum-frequency channel when all perturbation photons are upconverted (yellow line, obtained with $\epsilon_2 = 0$, $\gamma t = \pi/2$, where $t$ is the interaction time, $\langle n \rangle_0 (t = 0) = 20$).

**Extended Data**

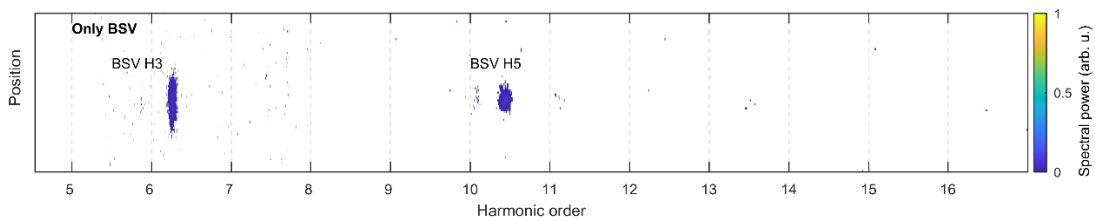

**Extended Data Figure 1 | Direct up-conversion of BSV.** When the BSV pulse energy is 30 nJ, the 3$^{rd}$ and 5$^{th}$ harmonics of the BSV frequency are measurable. Note that the spectral width is much narrower than the sidebands because the BSV bandwidth is limited to 50 nm, corresponding to a coherence time of 75 fs.

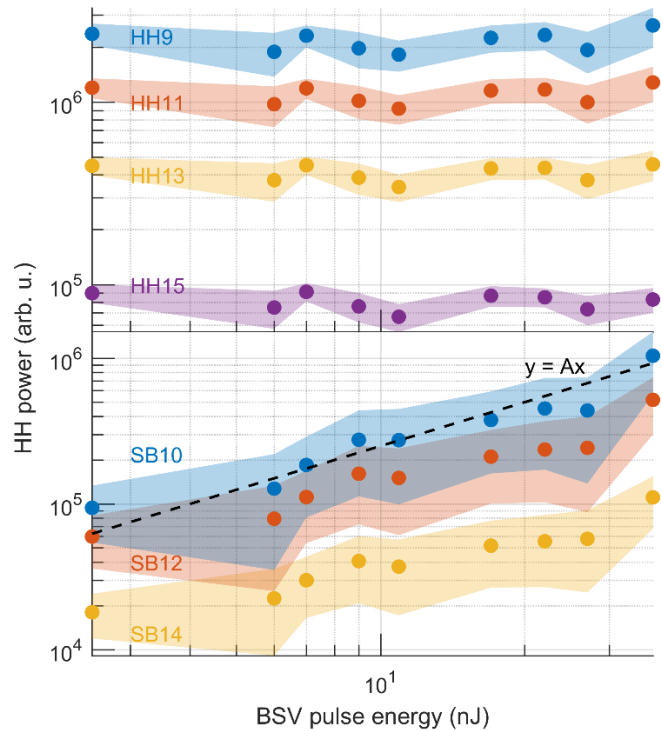

**Extended Data 2 | Scaling of sidebands and harmonics with BSV power.** The sideband's power scales linearly with the BSV pulse energy, in agreement with a wave-mixing interpretation of *in-situ* high-harmonic control, suggesting a one-photon mixing process in the BSV beam. The high harmonics remain unaffected up to 35 nJ of BSV.

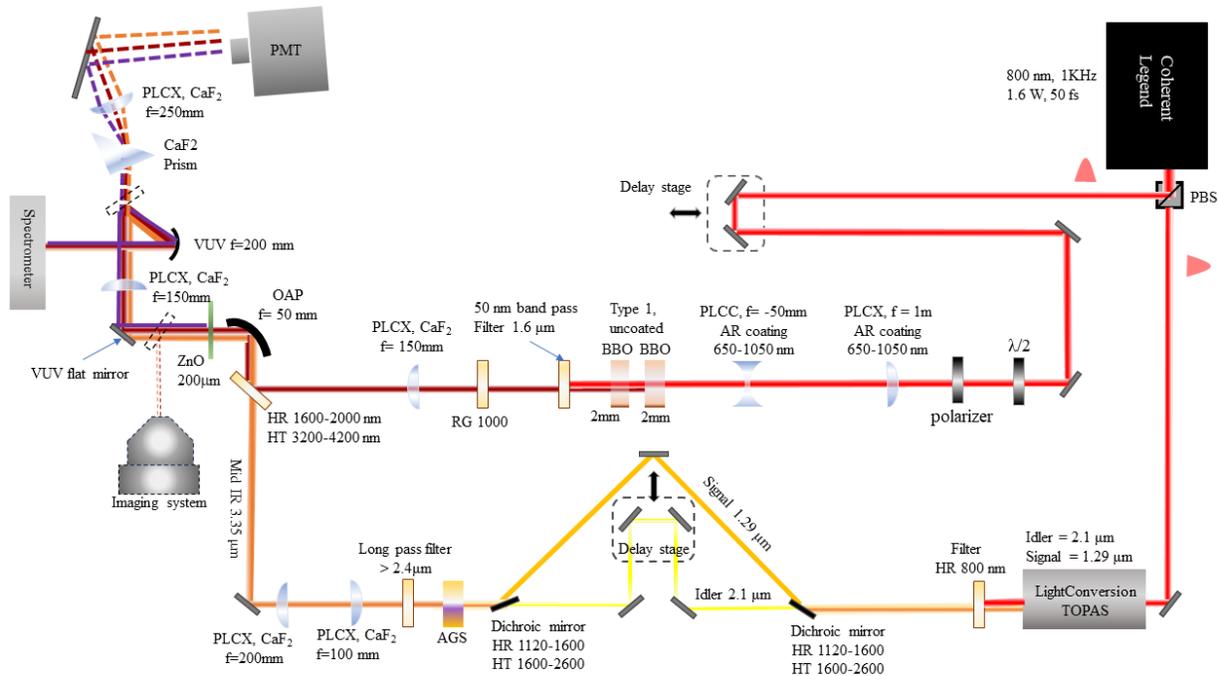

**Extended Data 3 | Sketch of experimental layout.** "PBS": Polarizing beam splitter, a combination of half-wave plate and thin-film polarizer; "PMT": Hamamatsu H9305-01 photomultiplier tube; "Spectrometer": Princeton Instruments IsoPlane 320, equipped with PI-MAX4 intensified camera; "OAP": Off-Axis Parabola, f= 25mm (Thorlabs MPD119-M01); "AGS": 400 μm-thick, θ = 39 deg, φ = 45 deg, AR coated 1 – 2.6 μm on the front surface, AR 2.6 – 11 μm on the back surface. "Imaging system" includes an InGaAs camera (Xenics Bobcat+ 320). All mirrors after the ZnO crystal are Al-coated vacuum ultraviolet mirrors (Acton coating #1900). Monitoring photodiodes for the mid-infrared and BSV beams are not shown.